\documentclass[12pt]{article}
\usepackage{cite}
\usepackage{fullpage}
\usepackage{graphicx}
\newcommand{\onlinecite}{\cite}

\title{Sitting at the edge: How biomolecules use \\ hydrophobicity to
tune their interactions and function}
\author{Amish J. Patel$^{1}$, Patrick Varilly$^{2}$, Sumanth
N. Jamadagni$^{1}$, \\ Michael F. Hagan$^3$, David Chandler$^{2,*}$, \& Shekhar Garde$^{1,*}$
\\
\normalsize{$^{1}$Howard P. Isermann Department of Chemical \& Biological Engineering,}\\\normalsize{and Center for Biotechnology \& Interdisciplinary Studies,}\\\normalsize{Rensselaer Polytechnic Institute, Troy, NY, 12180, USA.}\\
\normalsize{$^{2}$Department of Chemistry, University of California, Berkeley, CA, 94720, USA.}\\
\normalsize{$^{3}$Martin A. Fisher School of Physics, Brandeis
University, Waltham, MA 02454, USA.}\\
\normalsize{$^*$To whom correspondence should be addressed: gardes@rpi.edu, chandler@berkeley.edu}
}
\date{}

\begin{document} 
\maketitle 
\begin{abstract}
Water near hydrophobic surfaces is like that at a liquid-vapor interface, where fluctuations in water density are substantially enhanced compared to that in bulk water. 
Here we use molecular simulations with specialized sampling techniques to show that water density fluctuations are similarly enhanced, even near  hydrophobic surfaces of complex biomolecules, situating them at the edge of a dewetting transition. 
Consequently, water near these surfaces is sensitive to subtle changes in surface conformation, topology, and chemistry, any of which can tip the balance towards or away from the wet state, and thus significantly alter biomolecular interactions and function. Our work also resolves the long-standing puzzle of why some biological surfaces dewet and other seemingly similar surfaces do not.
\end{abstract}

Much of biology happens in aqueous environments, with interfaces of 
biomolecules being wet~\cite{Rossky:Nature:98,Berne:bphc}. Yet, 
hydrophobically driven assembly leads to contact surfaces that contain 
little or no water~\cite{Goodsell:Structure,Janin:Proteins}. 
How biomolecules are able to perform the task of removing water from their 
vicinity is an important open question~\cite{shea08}. 
Here, we show that the answer lies in the proximity of water near hydrophobic surfaces to an underlying phase transition.
Thermodynamically, water at ambient conditions is already close to the liquid-vapor phase boundary. 
Near a hydrophobic surface, water molecules are pulled away from the surface, because they interact only weakly with the surface, but have strong hydrogen bonding interactions with water in the bulk.
This further destabilizes water near hydrophobic surfaces and pushes it close to the edge of a dewetting transition.
Indeed, idealized repulsive hydrophobic surfaces nucleate a soft liquid-vapor like interface~\cite{FHS:1973,LCW,DC_nature05}.
However, weak attractive van der Waals forces exerted by realistic hydrophobic surfaces pull the soft interface closer, rewetting the surface, and masking the proximity to the underlying dewetting transition~\cite{HC,Rossky:JCP:1984,Godawat:PNAS,Mezger:PNAS:2006}. 
This proximity is then revealed not by the mean water density near the surface, but instead by fluctuations away from the mean~\cite{Patel:JPCB:2010}, and by the response of water density to perturbations~\cite{HC,DC_nature05}. 

Here we employ specialized sampling techniques~\cite{Patel:JSP:2011,Patel:JPCB:2010} to measure water density fluctuations near complex biomolecular surfaces, and show that they can be enhanced 
near a sufficiently hydrophobic patch.
Further, we demonstrate the sensitivity of water density to perturbations near such surfaces.
It is by exploiting this sensitivity to perturbations, {\it e.g.}, through subtle 
conformational changes, that biomolecules are able to dewet their surfaces.  
This near-the-edge behavior can also be important in the function of certain biomolecules, such as in the vapor-lock gating of ion channels.
Our results also explain why the gap between two hydrophobic protein surfaces was found to be wet in one case~\cite{Berne:bphc}, and dry in another~\cite{Berne:melittin}: the two systems are close to, but on either side of the dewetting transition. 

To illustrate that water density fluctuations provide a clear signature of
hydrophobicity, we characterize them near self-assembled monolayer (SAM)
surfaces with -CH$_3$ and -OH head groups, and in bulk
water~(Fig.~1A). Specifically, we calculate the probability distribution,
$P_v(N)$, of observing $N$ water molecules in a volume, $v$, of interest
(see Supplementary Information below).
The average number, $\langle N \rangle$, of water molecules in $v$, reflected in the peak of $P_v(N)$, is similar in all three cases, indicating that both the -CH$_3$ and the -OH SAMs are wet.
The density fluctuations in bulk water and at the hydrophilic -OH SAM are also similar and approximately Gaussian (parabolic on a log scale), as expected~\cite{Patel:JPCB:2010}. 
However, fluctuations near the hydrophobic -CH$_3$ SAM are different, with $P_v(N)$ enhanced significantly for low $N$-values~\cite{Patel:JPCB:2010}. 
Because the differences in the distributions at low $N$ correspond to fluctuations that are extremely rare, {\it i.e.}, $P_v(N)$ is very small, they do not affect the average equilibrium behavior.
However, these rare fluctuations play an important role in the presence of a perturbation.  
For example, the response of water near the -CH$_3$ and -OH surfaces to an unfavorable linear potential, $\phi N$, is very different.
Near the -OH surface, increasing $\phi$ shifts the distribution to the left~(Fig.~1C), and lowers the mean density of water, $\langle N\rangle_{\phi}$, gradually~(Fig.~1E).  
In contrast, near the -CH$_3$ surface, even small $\phi$-values, of the order of the thermal energy ($k_{\rm B}T$), are sufficient to dramatically alter the distribution, with low-$N$ fluctuations being enhanced so 
much that they become the most probable ones (Fig.~1D). 
As a result, even a small perturbation is able to dry the hydrophobic -CH$_3$ surface, and $\langle N\rangle_{\phi}$ decreases precipitously~(Fig.~1E). Correspondingly, the sensitivity of water density to perturbations, quantified by the susceptibility, $\partial\langle N\rangle_{\phi}/\partial\phi$, displays a peak near the hydrophobic surface (Fig.~1F); a known feature of phase transitions. 
Thus, the remarkable sensitivity of water density to perturbations is directly 
connected to the enhanced low-$N$ fluctuations, with both observations resulting 
from the proximity of water to an underlying dewetting transition~\cite{LCW,HC,DC_nature05,Patel:JPCB:2010}.

To explore whether the principles governing the dewetting of
hydrophobic interfaces illustrated above extend to the topologically and
chemically complex surfaces of biomolecules, we study a subunit of the
enzyme BphC, which includes a large hydrophobic patch at the boundary
between its two domains~\cite{Berne:bphc}. The protein surface is
rugged (Fig.~2A), which makes the calculation of $P_v(N)$ in volumes
that complement the protein surface challenging. To this end, we
developed an extension of the indirect umbrella sampling (INDUS) 
method~\cite{Patel:JSP:2011} that
enables the sampling of $P_v(N)$ in arbitrarily shaped
volumes. Fig.~2B shows that $P_v(N)$ near a hydrophilic patch is
bulk-like, similar to that near the -OH SAM.  In contrast, despite the
presence of a highly polar backbone and possible interactions with
charged side chains and counter-ions, $P_v(N)$ near a hydrophobic
patch on the isolated domain I surface shows enhanced low-$N$ 
fluctuations (Fig.~2C), similar to that demonstrated for the -CH$_3$ SAM.

To investigate how water near the hydrophobic patch responds to realistic 
perturbations, we place domain~II  near domain~I at different separations, 
$\Delta z$, and calculate $P_v(N)$ in the space between the two domains. 
As the domains are brought closer, while the likelihood of drying [$P_v(N\approx0)]$ increases~(Fig.~2D), the inter-domain region remains wet.
Even at $\Delta z = 0.4$ nm, where only a single layer of water
can be accommodated between the domains, the average value of $N$~
[peak of $P_v(N)$] is high, consistent with the findings of Ref.~\onlinecite{Berne:bphc}.   
This might seem surprising given the prevalence of hydrophobic 
residues on the protein surfaces surrounding the inter-domain region.
However, $P_v(N)$ distributions at all $\Delta z$ display enhanced low-$N$ fluctuations, suggesting that water is at the edge of a dewetting transition, albeit on the wet side. 
If this is indeed the case, a sufficient additional perturbation should be able
to push it over the edge and trigger dewetting.  
To test this, and motivated by Zhou {\it et al.}~\cite{Berne:bphc}, we turned off the
partial charges on the protein, which is expected to 
be an unfavorable perturbation, and disfavor the presence of water.
As a result, the hydrophobicity of the patch on the isolated domain I
surface is enhanced, as reflected in the enhanced fluctuations in
Fig.~2C. The effect of this perturbation on water in the inter-domain
region is dramatic: an essentially dry state is now preferred even at
a distance of 0.6 nm between the domain surfaces (Fig.~2E).

In contrast to the assembly of BphC domains, that of melittin dimers
represents a system which sits on the dry side of the dewetting
transition~\cite{Berne:melittin}. 
We calculated $P_v(N)$ near the putative hydrophobic face of
an isolated dimer surface (Fig.~3A), which indeed displays enhanced 
fluctuations~(Fig.~3B), as expected.
Correspondingly, when the system is perturbed by placing a second
wild-type dimer nearby, a drying transition in the space between the dimers is clearly seen in the $P_v(N)$ distributions in Fig. 3C. 
For $\Delta z = 0.9$ nm, the average value of $N$ [peak of $P_v(N)$] is high, implying a wet state; whereas for $\Delta z = 0.6$ nm, the space between the dimers is dry.
Interestingly, for intermediate $\Delta z = 0.7, 0.8$ nm, the $P_v(N)$ distributions  show bimodal behavior, indicating the presence of a desolvation barrier separating the wet (high $N$) and the dry (low $N$) states. 
Barriers along this simple order parameter, $N$, can play an important role in governing the kinetics of dimer-dimer assembly.

Although a drying transition is observed for the wild-type melittin
dimers, once again, it is extremely sensitive to small perturbations, such
as point mutations, as noted by Berne and co-workers~\cite{Berne:melittin}.
We calculated $P_v(N)$ distributions near six isolated melittin dimer mutants,
which all show enhanced fluctuations, with subtle differences between
them (see Fig.~S1).  
To highlight the effect of a perturbation on the drying transition in melittin, we selected a point mutant, Ile20Gly (Fig.~3A). There are only subtle differences in the $P_v(N)$ distributions near the isolated wild-type and the mutant dimer~(Fig.~3B).
However, they translate into dramatic differences in confinement, where the system is poised at the edge of a dewetting transition. 
Specifically, the Ile20Gly mutant no longer displays a drying transition, as the average value of $N$ remains high even at the smallest $\Delta z$~(Fig.~3D).  Thus, the data in Figs.~2 and 3 underscore that both BphC and melittin sit near the edge of a wetting-dewetting transition, one on the wet and the other on the dry side, and each can be pushed to the other side by a small perturbation.
This proximity to the phase transition is not evident in the mean behavior of water, but is revealed by water density fluctuations.

To enable regulation, biological systems are generically thought to position
themselves near phase transitions~\cite{Bialek:JSP:2011}. 
Our application of specialized techniques that
measure rare fluctuations allowed us to highlight that water near
hydrophobic surfaces of biomolecules is similarly situated at the edge
of a dewetting transition, and is sensitive to small
perturbations. This sensitivity provides biomolecules with the
powerful ability to tune their interactions and function by
manipulating the local context, for example, by confining water
between them, or by changing their shape or chemistry~\cite{Beckstein:PhysBiol:2004,Giovambattista:PNAS:08,Acharya:Faraday}.
Given that desolvation is a component of the reaction coordinate for
hydrophobically driven assembly~\cite{tWC}, manipulating
wetting-dewetting may appreciably influence the kinetics of
assembly. Dewetting transitions are also central in the function of some 
ion channels ({\it e.g.}, mechano sensitive channel MscS), where a 
10-20~\AA\ long interior hydrophobic wall of the channel provides gating by the
``vapor lock" mechanism -- a wet channel conducts ions rapidly;
whereas, a small conformational change can dry the channel and stop ion conduction completely~\cite{Anishkin1,Pomes:BJ,Hummer:PNAS}. 
Dewetting can also be induced by manipulating solution conditions ({\it e.g.}, 
temperature, pressure, pH, and co-solvent or solute concentration)~\cite{Berne:JPCC,Hummer:ARPC:08,Giovambattista:PNAS:08}.  
Finally, the sensitivity of nanoscale dewetting transitions 
is also responsible for the spontaneous filling and emptying of hydrophobic 
nanotubes~\cite{Hummer:Nature}, and can be harnessed in various other non-
biological settings, ranging from switches in nanofluidic devices and 
networks~\cite{Hummer:ARPC:08}, to aqueous solution based catalysis in 
hydrophobic zeolites~\cite{Davis:PNAS}.

\section*{Acknowledgments}
SG acknowledges partial financial support from NSF (CBET-0967937) and NSEC (DMR-0642573) grants. DC acknowledges financial support from NIH (R01-GM078102-04) grant. MFH acknowledges financial support from NIH (NIAID R01-AI080791) grant. We also thank Hari Acharya, Ravi Kane, and Kafui Tay for helpful discussions.

\begin{figure}
\begin{center}
\includegraphics[angle=0,width=5.4in]{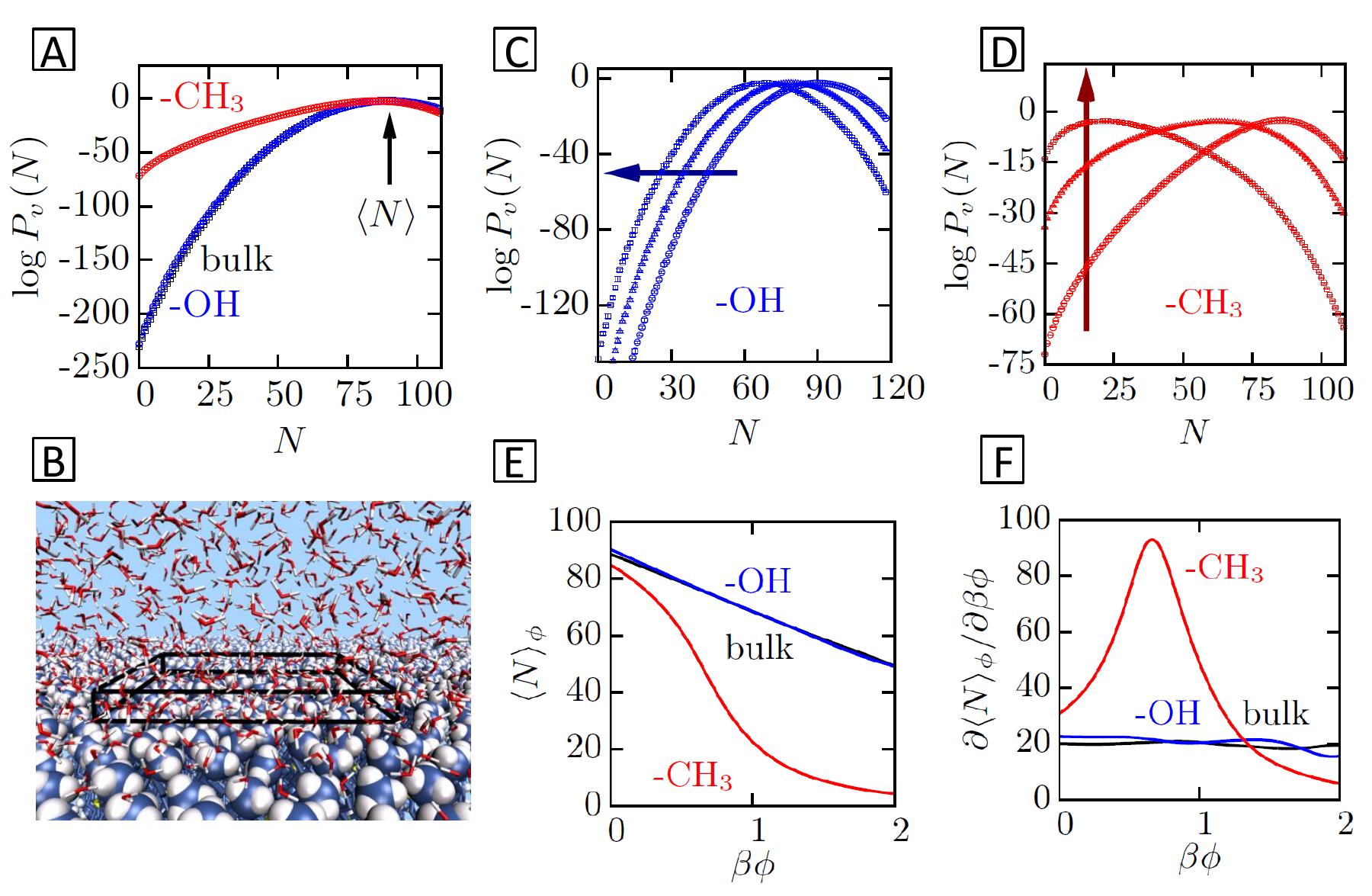}
\end{center}
\caption{Water density fluctuations and its response to
perturbations near hydrophobic and hydrophilic interfaces. (A) $P_v(N)$ in a 3 nm $\times$ 3 nm $\times$ 0.3 nm cuboid in bulk water
and at the interface of a self-assembled monolayer (SAM) with
hydrophobic (-CH$_3$) and hydrophilic (-OH) head groups. (B) A
close-up view of the hydrophobic -CH$_3$ SAM-water interface. Water
[sticks, red (oxygen) and white (hydrogen)], -CH$_3$ head groups
[spacefill, purple (carbon), and white (hydrogen)], and the cuboid
observation volume (black wireframe) are shown. (C) and (D) $P_v(N)$ distributions near the -OH and the -CH$_3$ SAMs, respectively, in the presence of an unfavorable linear potential, $\phi N$, for $\phi = 0$ (no perturbation), and $\phi =0.5~k_{\rm B}T$ and $1~k_{\rm B}T$. The arrow points in the direction of increasing $\phi$. (E) The response of the average number of water molecules in the cuboid volume, $\langle N \rangle_\phi$, to the external potential $\phi$, and (F) the corresponding susceptibility, $\partial \langle N
\rangle_\phi / \partial \beta\phi$, show signatures of a nanoscopic
phase transition near the -CH$_3$ surface. Error-bars calculated
using six separate simulation blocks are smaller than the symbols
used.}
\end{figure}

\begin{figure}
\begin{center}
\includegraphics[angle=0,width=5.4in]{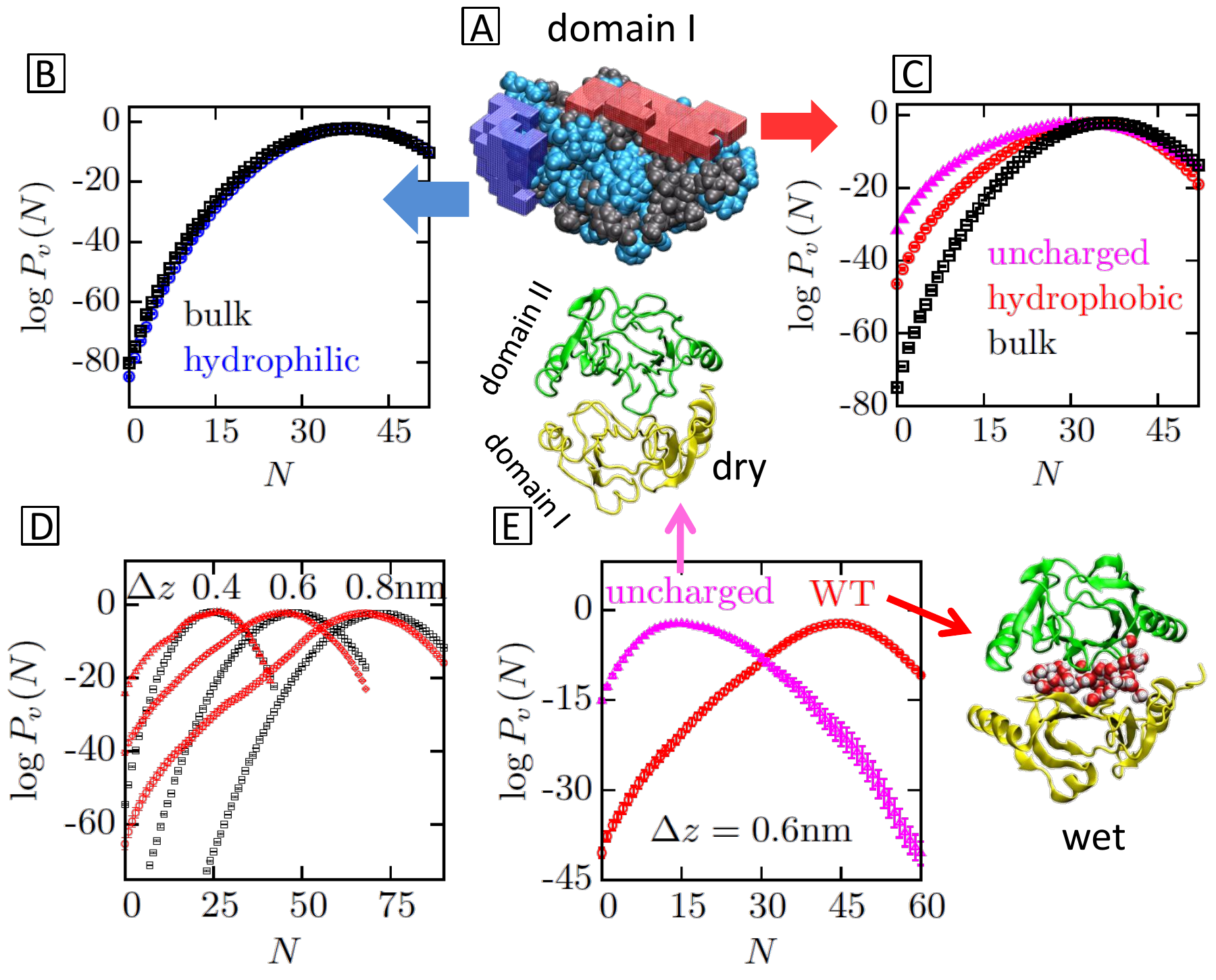}
\end{center}
\caption{Density fluctuations and the corresponding wetting and drying of BphC domains. (A) A snapshot of domain I (residues 1-135) of a BphC subunit (PDB: 1DHY) showing hydrophobic (gray) and hydrophilic (blue) regions.  Two separate 0.3 nm thick observation volumes complementing the protein surface near a hydrophobic (red) and a hydrophilic patch (blue), are also shown. (B) $P_v(N)$
near the hydrophilic patch is similar to that in bulk water. (C) $P_v(N)$ near the hydrophobic patch displays enhanced low-$N$ fluctuations. When electrostatic interactions between the protein and the water are turned off, the fluctuations are enhanced further, indicating that the protein surface becomes more hydrophobic. (D) $P_v(N)$ distributions in an observation volume
sandwiched between the two domains of the BphC subunit for different inter-domain separations, $\Delta z =$ 0.8,~0.6,~and~0.4~nm
(red). Distributions in similar volumes in bulk water are shown for
comparison (black). (E) $P_v(N)$ distributions in the inter-domain region for
$\Delta z =$~0.6~nm, for proteins with charges on [same as in (D)] and
off. Water and protein were represented using the TIP3P model and the AMBER-94 force field, respectively. Error-bars were calculated using six separate simulation blocks.}
\end{figure}

\begin{figure}
\begin{center}
\includegraphics[angle=0,width=5.4in]{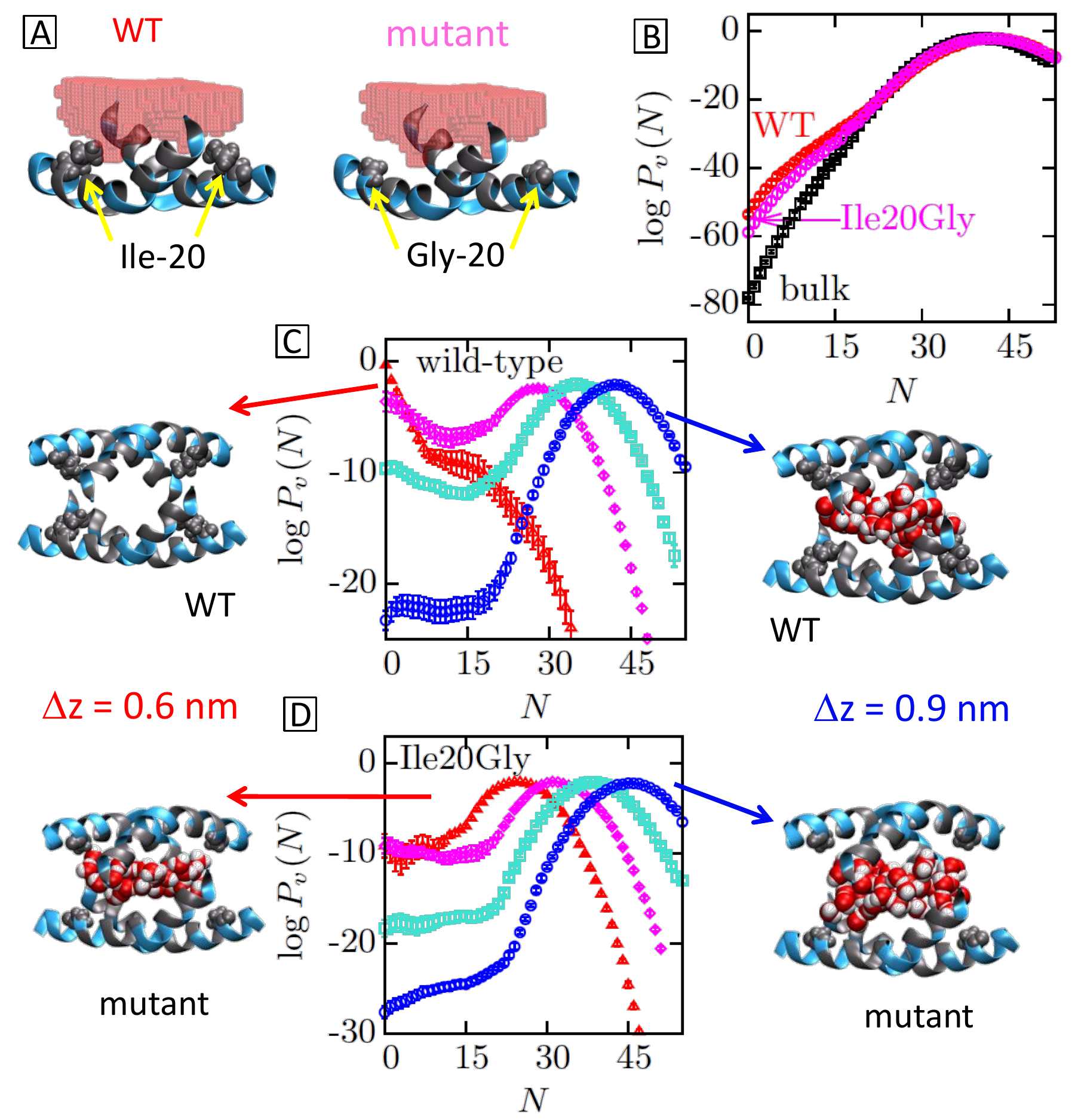}
\end{center}
\caption{Melittin dimer association -- sitting on the dry side. (A) A snapshot highlighting the observation volume near the hydrophobic surface of the wild type melittin dimer (PDB: 2MLT) and the Ile20Gly mutant dimer. The mutation is shown in spacefill representation. (B) $P_v(N)$ distributions in the observation volumes shown in (A) and in a similar volume in bulk water. (C) $P_v(N)$ distributions in the observation volume between two WT melittin dimers for dimer-dimer separations, $\Delta z$, of 0.9 (blue), 0.8 (cyan), 0.7 (magenta), and 0.6 nm (red), indicate a dewetting transition for $\Delta z < 0.7$ nm. Representative snapshots of the tetramer at $\Delta z$ = 0.6 (dry) and 0.9 nm (wet) are shown.  (D) Same as in (C) for the Ile20Gly mutants, where no dewetting is observed. Water and protein were represented using the SPC/E model and the AMBER-99sb force field, respectively. Error-bars were calculated using six separate simulation blocks.}
\end{figure}

\newpage

\section*{Supplementary Information}

\subsection*{Materials and Methods}

All simulations were performed using the {\small GROMACS} 
molecular dynamics package, suitably modified to allow for importance 
sampling in the NVT ensemble ($T=300$~K) with a buffering vapor-liquid 
interface. 
Here, we provide further details of the BphC and melittin
simulations. We also describe how we select probe volumes to monitor
water density fluctuations near proteins.\\

\noindent
{\bf BphC:} Domain I of BphC was solvated in $\sim7500$ TIP3P water
molecules, whereas simulations of both domains I and II contained
$\sim9500$ waters. To study the region between the two domains, domain
II was translated along the vector joining the center of masses of the
two domains by 0.4, 0.6 and 0.8 nm from the crystal structure (PDB:
1DHY). 21 atoms in each domain of BphC were position restrained
harmonically. Bonds involving hydrogen atoms were constrained using
the SHAKE algorithm, and temperature was maintained at 300K using the
Berendsen thermostat.

To select a probe volume near a hydrophobic patch in the hydration
shell of the protein (Fig. 2a), a cubic grid with a spacing of 0.3 nm
was placed in the simulation box.  A sufficiently large, contiguous
patch of $\sim40$ grid cells was then chosen as the probe volume, such
that (i) no cell contained protein heavy atoms, (ii) each cell was
within 0.4 nm of at least one of the designated hydrophobic amino
acids (Val, Leu, Ile, and Phe), and (iii) each cell was at least 0.8
nm away from the designated hydrophilic amino acids (Lys, Arg, Asp,
and Glu).  The probe volume near a hydrophilic patch was chosen
similarly (Fig. 2a).  The probe volumes in confined systems were
chosen with the additional constraint that the cells had to be in the
region between the two domains (Fig. 2d and 2e).\\

\noindent
{\bf Melittin:} The melittin dimer as well as the tetramer were
solvated in $\sim5000$ SPC/E water molecules. To study the region
between the two dimers, one of the dimers from the tetramer crystal
structure (PDB: 2MLT) was translated along the dimer-dimer vector by
0.6, 0.7, 0.8 and 0.9 nm. The backbone atoms of the dimers were
constrained harmonically. Bonds within melittin were constrained using
the P-LINCS algorithm, while those in the water were constrained with
the SETTLE algorithm.  Temperature was maintained at 300K using the
canonical velocity rescaling thermostat by Bussi and co-workers.

To select the probe volume for a given separation between the dimers,
the convex hull of the C$^\alpha$ positions of residues $8$~and~$20$
of each monomer ({\it i.e.}, 8 points in total) was first calculated,
and a 0.1 nm-resolution grid was then placed in the system.  Grid
cells whose center was (i) inside the convex hull and (ii) outside the
van der Waals radius of each protein heavy atom were selected to be
part of the probe volume. The probe volume thus constructed was
slightly different for each mutant. To select a probe volume near an
isolated dimer, the convex hull for separation of 0.4 nm between the
wild-type dimers was computed. Grid cells in this convex hull that do
not overlap with protein heavy atoms of one of the dimers constitute
the probe volume of that dimer. The same probe volume was used for all
mutant dimers, allowing a direct comparison of density fluctuations
near them.

Additionally, in all BphC and melittin simulations, electrostatic
interactions were calculated using the particle-mesh Ewald algorithm,
and the net charge on the protein was neutralized by adding the requisite
numbers of Na$^+$ or Cl$^-$ ions.

\newpage

\subsection*{Supplementary Figure}

\begin{figure}[h]
\centering
\includegraphics[width=6in]{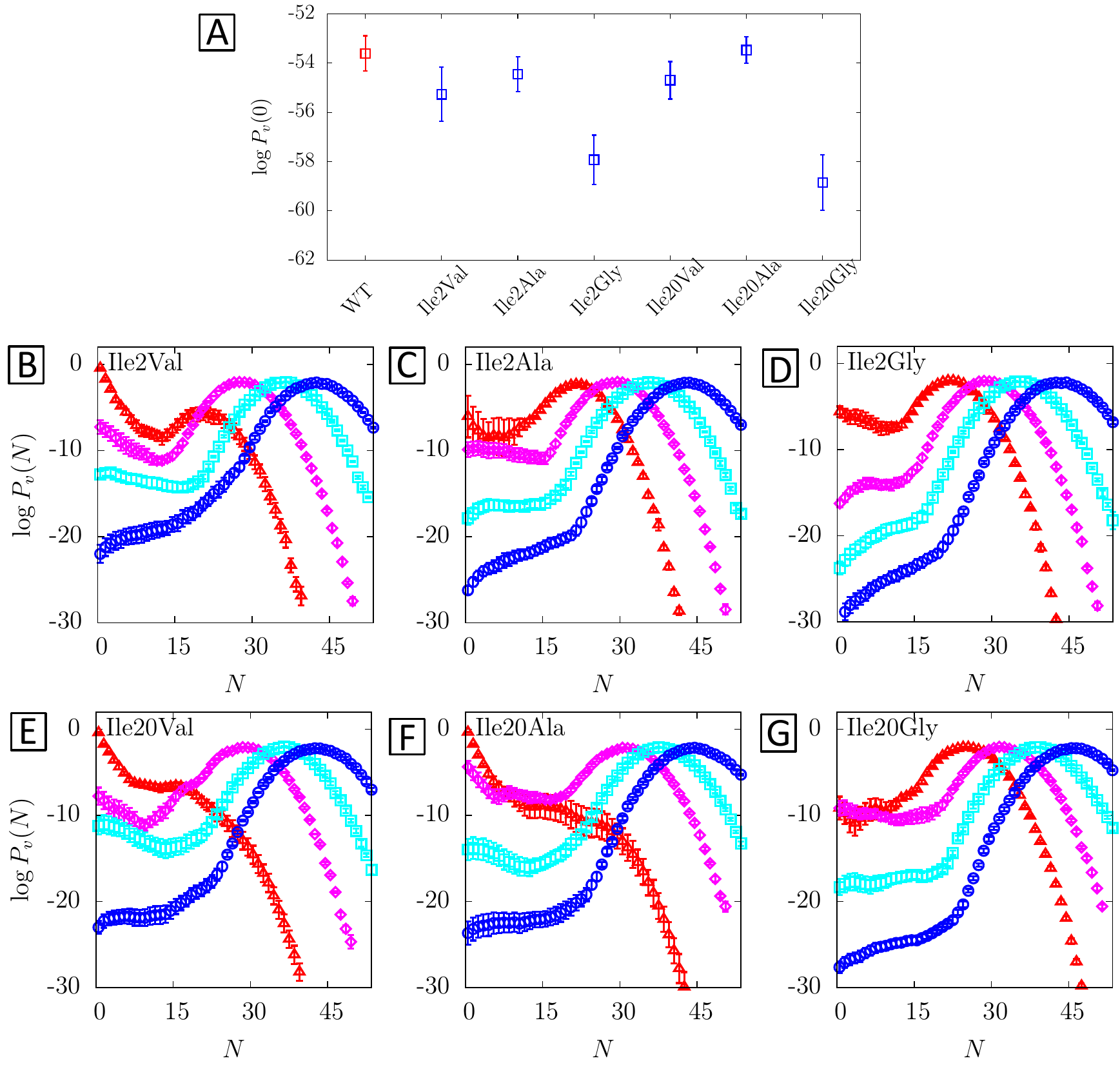}
\end{figure}
\noindent
{\bf Figure S1.} Water density fluctuations for mutant
melittin dimers and tetramers. (A) $\log P_v(0)$ for probe volume $v$
near six mutant melittin dimers, compared to that near the wild-type
dimer. Mutating the more hydrophobic Ile residue (at position 2 or 20)
in the wild-type dimer to Val, Ala, and Gly reduces the value of $\log
P_v(0)$. (B), (C), and (D) show $\log P_v(N)$ for probe volume $v$
between mutant melittin dimers, Ile2Val, Ile2Ala, and Ile2Gly,
respectively, for dimer-dimer separations, $\Delta z$, of 0.9 (blue),
0.8 (cyan), 0.7 (magenta), and 0.6 nm (red).  Mutating Ile2 to Ala or
Gly destabilizes the dry state to the extent that the space between
dimers is wet even at $\Delta z = 0.6$ nm. (E), (F), and (G) show
similar data for mutations of Ile20. In this case, only the Ile20Gly
mutant is wet at $\Delta z=0.6$ nm.

\end{document}